\documentclass[traditionalabstract]{aa}

\usepackage{graphicx}
\usepackage{txfonts}
\usepackage[]{hyperref} 
\usepackage{caption}
\usepackage{float}
\usepackage[]{natbib}

\usepackage{xcolor}
\hypersetup
{
    colorlinks,
    linkcolor={red!50!black},
    citecolor={blue!70!black},
    urlcolor={blue!80!black}
}

\makeatletter
\renewcommand*\aa@pageof{, page \thepage{} of \pageref*{LastPage}}
\makeatother

\begin{document}

   \title{Confirmation of a metallicity spread  amongst first population stars in globular clusters}


   \author{Carmela Lardo\inst{1}\fnmsep\thanks{e-mail: carmela.lardo2@unibo.it}
          \and Maurizio Salaris\inst{2,3}
          \and Santi Cassisi\inst{3,4}
          \and Nate Bastian\inst{5,6}
          }

   \institute{Dipartimento di Fisica e Astronomia, Università degli Studi di Bologna, Via Gobetti 93/2, I-40129 Bologna, Italy
         \and
            Astrophysics Research Institute, Liverpool John Moores University, 146 Brownlow Hill, Liverpool L3 5RF, UK
        \and
        INAF - Osservatorio Astronomico di Abruzzo, Via M. Maggini, I-64100 Teramo, Italy
        \and
        INFN - Sezione di Pisa, Largo Pontecorvo 3, I-56127 Pisa, Italy
        \and
        Donostia International Physics Center (DIPC), Paseo Manuel de Lardizabal, 4, E-20018 Donostia-San Sebastián, Guipuzkoa, Spain
        \and
        IKERBASQUE, Basque Foundation for Science, E-48013 Bilbao, Spain
}

   \date{Received xxx XXX, xxx; accepted xxx, XXX}

 
  \abstract{Stars in massive star clusters exhibit intrinsic variations in some light elements (the multiple populations phenomenon) that are difficult to explain in a fully coherent formation scenario. In recent years, high quality {\em Hubble Space Telescope} (HST) photometry has led to the characterisation of the global properties of these multiple populations in an unparalleled level of detail. In particular, the colour-(pseudo)colour diagrams known as \lq{chromosome maps\rq} have been proven to be very efficient at separating cluster stars with field-like metal abundance distribution (first population) from object with distinctive light-element abundance anti-correlations (second population).
  
  The unexpected wide colour ranges covered by the first population group --traditionally considered to have a uniform chemical composition-- in the chromosome maps of the majority of the investigated Galactic globular clusters have been recently attributed to intrinsic metallicity variations up to $\sim$0.30~dex, from the study of subgiant branch stars in two metal rich Galactic globular clusters by employing appropriate $HST$ filter combinations. On the other hand, high-resolution spectroscopy of small samples of first populations stars in the globular clusters NGC~3201 and NGC~2808 --both displaying extended sequences of first population stars in their chromosome maps-- have so far provided conflicting results, with a spread of metal abundance detected in NGC~3201 but not in NGC~2808. 
  
  We present here a new method that employs $HST$ near-UV and optical photometry of red giant branch stars, to independently confirm these recent results. Our approach has been firstly validated using observational data for M~2, a globular cluster hosting a small group of first population stars with enhanced (by $\simeq$ 0.5~dex) metallicity with respect to the main component. We have then applied our method to three clusters that cover a much larger metallicity range, and have well populated, extended first population sequences in their chromosome maps, namely M~92, NGC~2808, and NGC~6362. We confirm that metallicity spreads are present among first population stars in these clusters, thus solidifying the case for the existence of unexpected variations up to a factor of two of metal abundances in most globular clusters. We also confirm the complex behaviour of the mean metallicity (and metallicity range) differences between first and second population stars.
}


\titlerunning{Metallicity spread in globular clusters P1 stars}
\authorrunning{Lardo et al.}
\maketitle

\section{Introduction} \label{sec:intro}

It has been now well established that Galactic globular clusters (GCs) host multiple populations (MPs) of stars, at odds with the traditional paradigm of star clusters harbouring objects all with the same age and uniform initial chemical composition.
A large number of spectroscopic observations have in fact shown that individual GCs are 
characterised by anti-cor\-re\-la\-ted star-to-star variations among C, N, O, Na (in some cases also Mg and Al) and He \citep[see, e.g.,][for reviews]{gcb:12, bl18, gratton:19, csreview20}, while 
their colour magnitude diagrams (CMDs) confirm negligible age spreads despite the chemical inhomogeneity.

The more popular scenarios for the formation of MPs invoke subsequent episodes of star formation \citep[see, e.g.][]{dercole08, decressin08, r22} whereby stars with CNONa (and He) abundance patterns 
similar to those observed in the 
field are the first stars to form (we will denote them as P1), while stars 
showing a range of N and Na (and He) enrichment and C and O depletion 
(we will denote them as P2) were formed several $10^6$ up to $\sim10^8$ years 
later --depending on the adopted scenario-- from chemically processed material, ejected by some class of more massive objects born in the first epoch of star formation.
All the proposed scenarios have however difficulties in quantitatively matching the 
observed abundance patterns, and no consensus has been reached yet on the mechanism responsible for the formation of MPs in a cluster \citep[see, e.g.][]{bastian15,r15,bl18}.

Complementary to high-resolution spectroscopy, photometry has been crucial to enlarge the sample of clusters investigated for the presence of MPs, the sample of stars surveyed
in each cluster, and the range of evolutionary phases (including the main sequence, generally too faint to be investigated spectroscopically in GCs) where chemical abundance variations have been detected \citep[see, e.g.,][]{sumo, p15, m17, nieder17, dondoglio}. Photometry has also been instrumental for the discovery and characterisation of MPs in massive extragalactic old- and intermediate-age clusters \citep[see, e.g.,][and references therein]{larsen14, dale16, gilligan19, h19,lagioia19,lagioiab, mart19, nardiello19, sara19, sara20, cadelano22}, demonstrating that the MP phenomenon is not restricted to the Milky Way oldest massive clusters.

The ability of photometry to detect MPs rests on the fact that abundance 
variations of the elements involved in the anticorrelations affect stellar effective temperatures, 
luminosities, and spectral energy distributions 
\citep[see, e.g.,][]{salaris:06, M4UBV, yong:08, sswc:11, cmpsf:13, dale16, mucc16, m17, dale18, s19}. For example, filters covering wavelengths shorter than $\lesssim$4500~\AA~can be especially sensitive to star-to-star differences in C, N, and O abundances. 


\begin{figure*}
\centering
\includegraphics[width=0.95\textwidth]{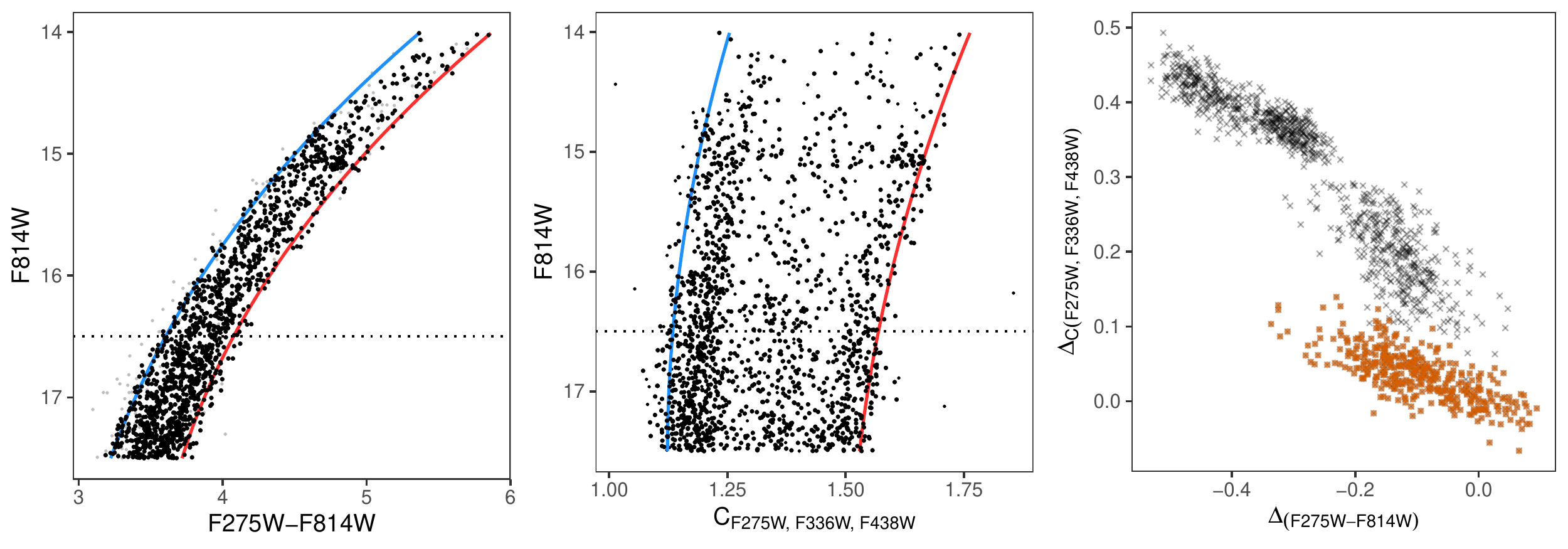}
\caption{Chromosome map of RGB stars in NGC~2808. The left and middle panel display, respectively, the $F814W$-$(F275W-F814W)$ CMD  
and the $F814W$-$C_{F275W, F336W, F438W}$ diagram, with overimposed the corresponding red and blue 
fiducial lines (see text for details). The right panel displays the resulting chromosome 
map. P1 stars are displayed as orange open circles, P2 stars are shown as black crosses.
\label{fig:chmap}}
\end{figure*}

Taking advantage of photometry in the {\em Hubble Space Telescope} ($HST$)  
Wide Field Camera 3 filters $F275W$, $F336W$, and $F438W$  
from the {\sl UV legacy survey of Galactic GCs} \citep[see, e.g.,][]{p15}, and data  in the $F814W$  filter from the Wide Field Channel of the $HST$ Advanced Camera for Survey \citep[][]{sarajedini}, \citet{milone:15, m17} have introduced the pseudo two-colour diagram $\Delta_{F275W, F814W}$-$\Delta_{C\ F275W, F336W, F438W}$ named \lq{chromosome map\rq}. In this diagram, different  populations can be easily identified, especially when considering red giant branch (RGB) stars. 
The RGB stars belonging to the P1 population of a cluster are expected to be generally distributed around the origin 
of the chromosome map coordinates ($\Delta_{F275W, F814W} \sim$ 0, $\Delta_{C\ F275W, F336W, F438W} \sim$ 0), 
covering a 
narrow range of $\Delta_{F275W, F814W}$  and $\Delta_{C\ F275W, F336W, F438W}$ values, whilst P2 stars (with a range of abundances of 
C, N, O, Na and He) span a wide range of both coordinates \citep{milone:15, m17, carretta18}. 
 
On the other hand, as reported by \citet{m17}, the large majority of their sample of 57 
Galactic GCs display $\Delta_{F275W, F814W}$  values (and $\Delta_{C\ F275W, F336W, F438W}$ to a much lesser extent) for P1 stars that cover a range much larger than what 
expected from photometric errors only. 
Unresolved binaries do certainly contribute to produce an extended P1 sequence in 
$\Delta_{F275W, F814W}$ , but their fraction in GCs is too small to fully explain 
the phenomenon, as shown by \citet{marino:19} and \citet{martins} -- see 
also the discussion in \citet{kamann}.

The reason for these extended P1 sequences must be therefore some unexpected level of chemical non-uniformity, that does not alter the relative abundances of the metals. Identifying the culprit of this inhomogeneity has therefore a major impact on our understanding of the mechanism of formation of globular clusters.

In the assumption that P1 stars in a cluster have a uniform metal content, \citet{milone:15}, \citet{He18} and \citet{lsb18} have shown that variations of the initial abundance of He can explain the extended sequences, because of their effect on the stellar effective temperatures: An increased initial He abundance produces hotter, hence bluer RGB stars at a given luminosity, and 
lower $\Delta_{F275W, F814W}$ values in the chromosome maps.
The analysis of GC chromosome maps by \citet{marino:19b} concluded more generally that the extended P1 sequences 
can be due either to a range of 
initial helium abundances, or to a range of metallicity (of about 0.1~dex), the more metal poor component populating the lower $\Delta_{F275W, F814W}$ values (hotter and bluer RGB stars).

The assumption of uniform metallicity was corroborated by the spectroscopy  
of six RGB stars distributed along the extended P1 of the GC NGC~2808 by \citet{cabrera19}, which did not reveal any significant spread in metallicity.
Also \citet{latour} did not find any metallicity variation among P1 stars in NGC~2808, by employing MUSE spectra for 1115 RGB stars distributed along 
the various sequences of the cluster chromosome map.
On the other hand, the spectroscopic study by 
\citet{marino:19} of 18 RGB stars belonging to the extended P1 of NGC~3201 has disclosed a 
[Fe/H] range (hence a range of total metallicity if the abundance ratios among the more 
abundant metals are uniform, as typically measured in P1 stars) on the order of 0.1-0.15~dex. 

Very recently, \citet{legnardi} have studied in detail two metal rich GCs with extended P1, 
namely NGC~6362 and NGC~6838 ([Fe/H]$\sim -1.10$ and $\sim -$0.8, respectively), devising appropriate combinations of magnitudes in the 
$F275W$, $F336W$, $F438W$ and $F814W$ filters for the clusters' P1 subgiant branch stars, 
to disentangle the effect of metallicity and helium variations. By comparisons with theoretical subgiant branch isochrones, they found that a range of metallicity and not $Y$ is present among P1 stars in these two clusters and, by extrapolation, in all other GCs with extended P1 in their chromosome maps. By employing the width of the P1 sequences in the $(F275W-F814W)$ colour 
at a reference 
$F814W$ magnitude determined by \citet{m17} for 55 GCs, \citet{legnardi} estimated from theoretical isochrones the presence of metallicity ranges --denoted in terms of [Fe/H] ranges-- ranging from a 
few 0.01~dex up to much higher values of 0.15-0.30~dex for metal poor GCs, at odds with inferences from spectroscopic analyses, that point to uniform 
[Fe/H] abundances in the majority of Galactic GCs \citep[see, e.g.,][]{carrettametal}.
This result has therefore major implications not only for the models of massive clusters' formation, but also for the spectroscopic investigations of chemical abundances in GCs. 


For this reasons we believe that a further investigation of the origin of the extended P1 sequences is warranted, to corroborate independently the results by \citet{legnardi}. To this purpose, we present here an alternative method to assess whether a 
metallicity or helium spread are the cause of the extended P1 in Galactic GCs. It makes use of information from photometry like \citet{legnardi} study but, instead of subgiants, it uses the same RGB stars along the extended P1 sequence of the chromosome maps to disentangle the effect of metallicity and $Y$, and we show here our results for the GCs NGC~6341 (M~92), NGC~2808, and NGC~6362 \citep[one of the two clusters investigated by ][with their method to disentangle the effect of metallicity and helium]{legnardi}, which cover a large metallicity range, between $\sim -$2.15 and $\sim-$1.10.

Section~\ref{method} introduces the method, based on the properties of theoretical isochrones tested 
empirically on the GC M~2, and Sect.~\ref{example} presents an application to our sample of three clusters. Section~\ref{discussion} closes the paper by discussing the results.

\section{RGB stars in the $F814W$-$(F275W-F814W)$ colour magnitude diagram}
\label{method}

The photometric identification of P1 RGB stars takes advantage of  
the chromosome maps described by \citet{m17}, that make use of the 
combination of a $F814W$-$(F275W-F814W)$ CMD  
and a $F814W$-$C_{F275W, F336W, F438W}$ diagram, where the pseudocolor 
$C_{F275W, F336W, F438W}$ is defined as $C_{F275W, F336W, F438W}=(F275W-F336W)-(F336W-F438W)$.
They are produced as follows: 
First, red and blue fiducial lines in each of these two diagrams are calculated by determining the values of the 4th and the 96th percentile of the $(F275W-F814W)$ and $C_{F275W, F336W, F438W}$ distributions
in various magnitude bins across the RGB, as shown in the left and middle panel of Fig.~\ref{fig:chmap} for the cluster NGC~2808 \citep[the photometry for this cluster and all other clusters discussed in this paper is from][]{ndatabase}.
In these diagrams P2 stars are 
distributed between the red and blue fiducials, whilst the supposedly chemically homogeneous P1 stars are expected to lie around the red fiducials.
As a second step, the width of the RGB in the $(F275W-F814W)$ colour
(denoted here as $W_{F275W,F814W}$) 
and in the $C_{F275W, F33W, F438W}$ pseudocolour
(denoted as $W_{C\ F275W,F336W,F438W}$) 
is calculated at a reference luminosity, 
set to two $F814W$ magnitudes above the cluster turn off, taking the colour and pseudocolor differences between the red and blue fiducial.
Finally, for each RGB star the differences in $(F275W-F814W)$  and 
$C_{F275W,F336W,F438W}$ with respect to the red fiducials taken at the star $F814W$ magnitude are calculated, and normalised to the values of $W_{F275W,F814W}$ and $W_{C\ F275W,F336W,F438W}$. 
These quantities are denoted as $\Delta_{F275W, F814W}$ and 
$\Delta_{C\ F275W, F336W, F438W}$ respectively, and are defined as follows:

\begin{equation}
\Delta _{F275W,F814W}= W_{F275W,F814W} \frac{X-X_{\rm R}}{X_{\rm R}-X_{\rm B}}
\label{eqx}
\end{equation}

\begin{equation}
\Delta _{C\ F275W,F336W,F438W} = W_{C\ F275W,F336W,F438W} \frac{Y_{\rm R}-Y}{Y_{\rm R}-Y_{\rm B}},
\label{eqy}
\end{equation}

where X=$(F275W-F814W)$, Y=$C_{F275W,F336W,F438W}$, and R and B denote the red and blue fiducial line, respectively.
The chromosome map is the plot of the stars' position in a $\Delta_{C\ F275W, F336W, F438W}$ vs $\Delta_{F275W, F814W}$ diagram, shown in the right panel 
of Fig.~\ref{fig:chmap} for the GC NGC~2808.

Following the definitions of these two quantities, the 
values $\Delta_{F275W, F814W}$=0 and 
$\Delta_{C\ F275W, F336W, F438W}$=0 
correspond to objects lying on the red fiducial
lines while $\Delta$ values different from zero denote colour and 
pseudocolour distances (defined as positive for $C_{F275W,F336W,F438W}$, and negative for $(F275W-F814W)$) from such lines.


The P2 stars typically follow a sequence extending from the origin of both coordinates towards increasing 
$\Delta_{C\ F275W, F336W, F438W}$ and decreasing $\Delta_{F275W, F814W}$, as shown in 
Fig.~\ref{fig:chmap}, whilst  
P1 stars should be clustered around the origin because of their expected chemical homogeneity, 
with a small spread just due to photometric errors \citep[see, e.g.][for more details]{m17, csreview20}.
However, as already mentioned, a large number of GCs display extended P1 
sequences in the chromosome map of their RGB stars, like the case of NGC~2808 shown in Fig.~\ref{fig:chmap}, where 
the sequence of P1 stars is clearly elongated mainly 
along the $\Delta_{F275W, F814W}$ axis \citep[see, e.g.][for more details about the identification 
of this elongated sequence with P1 stars]{m17, csreview20}, meaning that P1 objects describe multiple RGB sequences between the red and blue fiducials in the $F814W$-$(F275W-F814W)$ CMD. 
In the chromosome maps these multiple RGB sequences are compressed essentially along one segment, because the way the maps are calculated. In the following we show how we can employ the $F814W$-$(F275W-F814W)$ CMD of P1 stars to help discriminate between inhomogeneities in He and metallicity as the origin of the extended P1 phenomenon.

\begin{figure}
\centering
\includegraphics[width=0.90\columnwidth]{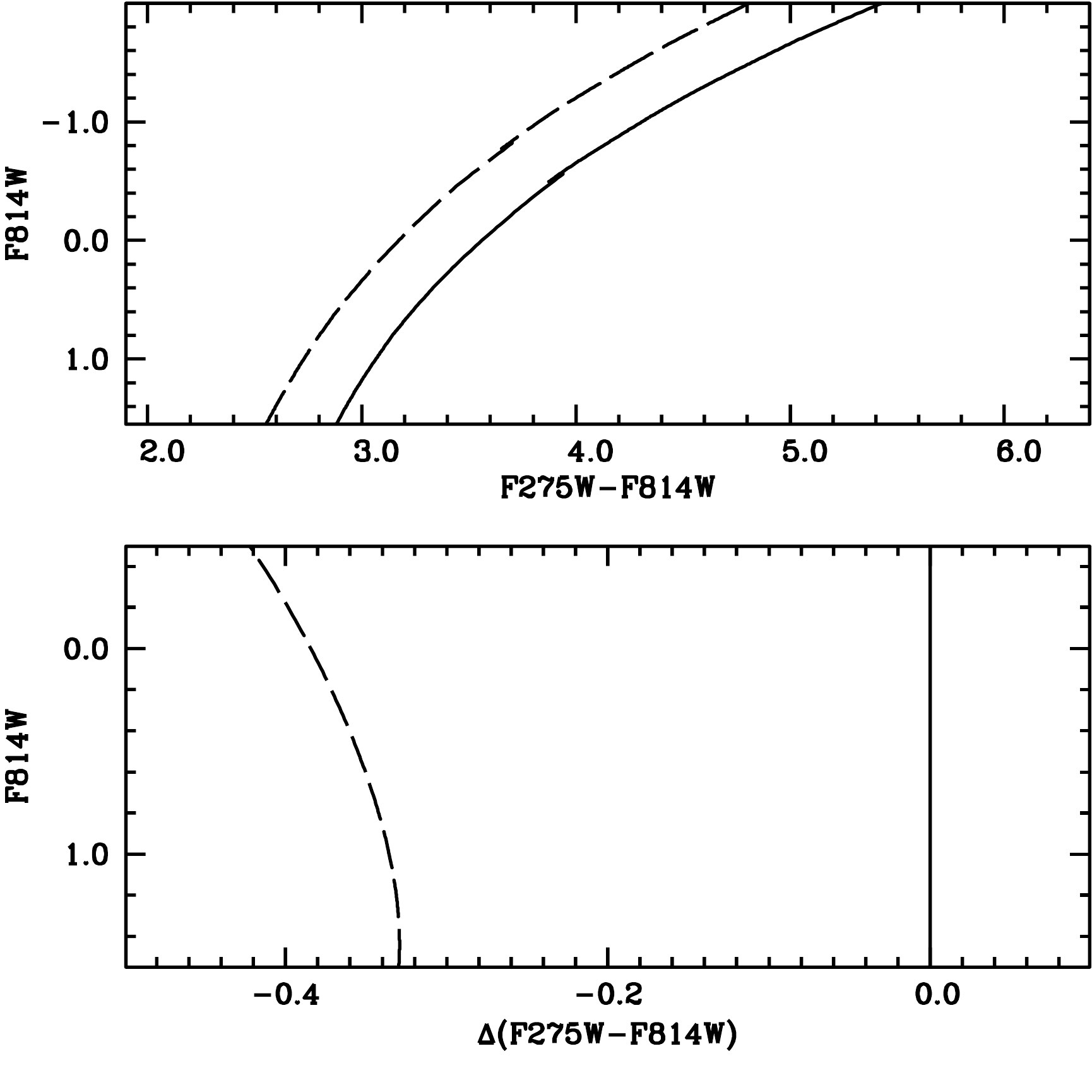}
\caption{{\sl Upper panel:} The $F814W$-$(F275W-F814W)$ CMD of 
two RGBs belonging to 12~Gyr old isochrones with [Fe/H]=$-$1.6 
(dashed line) and $-$1.3 (solid line), respectively. {\sl Lower panel:} Difference 
of the $(F275W-F814W)$ colour at varying $F814W$ between 
the [Fe/H]=$-$1.6 (dashed line) and the \lq{verticalised\rq} metal richer RGBs (solid line),  
taken as reference (see text for details). 
\label{fig:deltaiso}}
\end{figure}

Figure~\ref{fig:deltaiso} displays the $F814W$-$(F275W-F814W)$ CMD 
of two $\alpha$-enhanced 12~Gyr isochrones 
with [Fe/H]=$-$1.3 and $-$1.6 respectively, from 
\citet{pietrinferni:06}. The relative abundance distribution of the metals 
is typical of P1 stars and is the same for the two isochrones; it is just the total metallicity (hence [Fe/H]) that varies.
Due to the shape of the RGBs in this CMD, the colour difference at fixed magnitude 
between the two isochrones increases with 
decreasing $F814W$. This is more clearly seen in the lower panel of the figure, where we display the more metal rich RGB as a reference straight line at $(F275W-F814W)$=0, and plot 
on the horizontal axis the colour differences $\Delta(F275W-F814W)$ 
between the metal poorer RGB and this reference line at varying $F814W$, 
in the range between about 2 and 4 magnitudes above the main sequence turn off, which is approximately the magnitude range considered in the analysis presented in the next section.

\begin{figure}
\centering
\includegraphics[width=0.90\columnwidth]{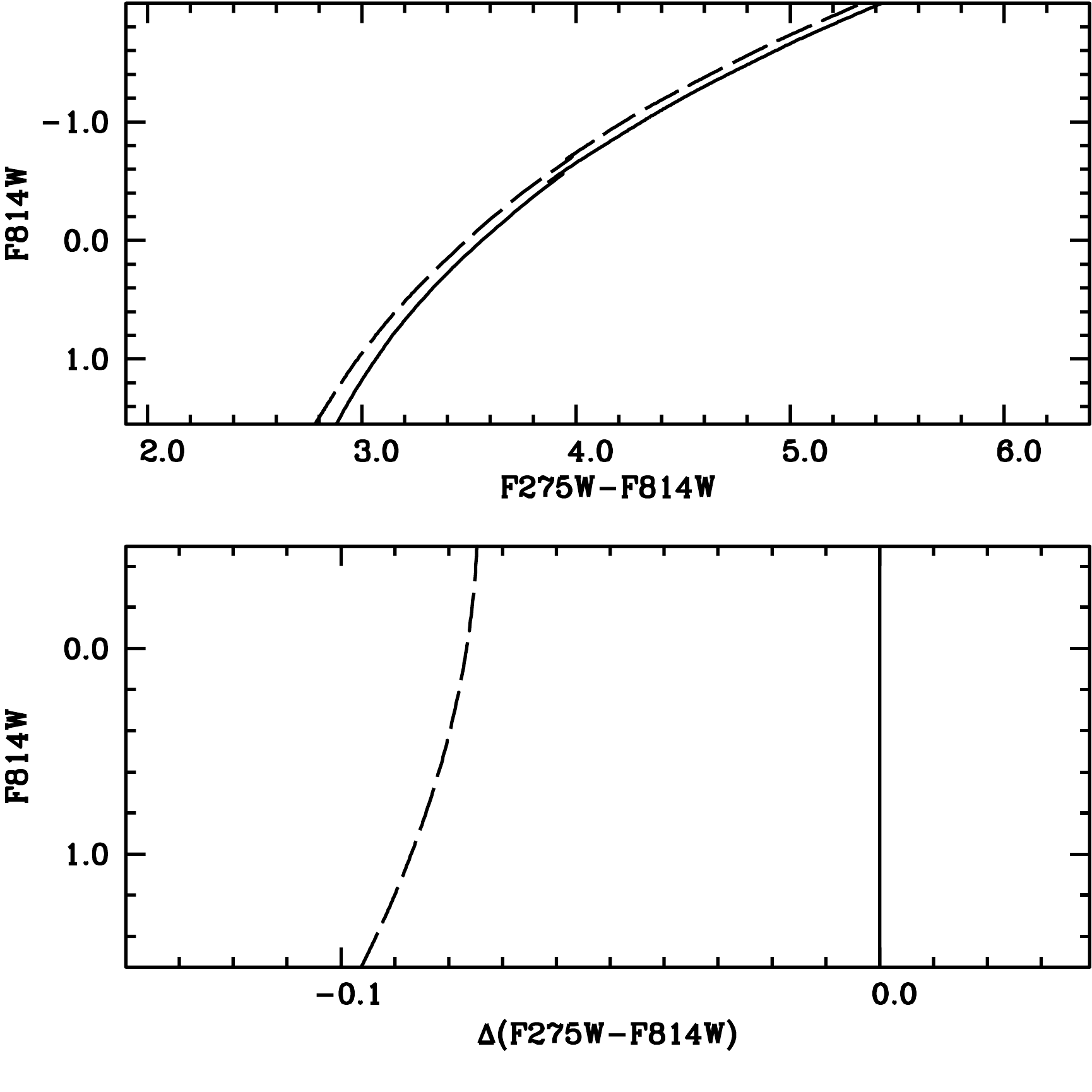}
\caption{As Fig.~\ref{fig:deltaiso} but this time the dashed line represents a 12~Gyr old isochrone with [Fe/H]=$-$1.3 and the helium mass fraction $Y$ increased by 
0.05.
\label{fig:deltaisoHe}}
\end{figure}

The colour separation $\Delta(F275W-F814W)$ increases when moving to brighter magnitudes, while 
the effect of a spread of initial helium abundance (at fixed metallicity) is different,  
as shown by Fig.~\ref{fig:deltaisoHe}. Here we display the RGB of the 
[Fe/H]=$-$1.3 isochrone of the previous figure, together with an isochrone for 
the same metallicity and age, but calculated with the helium mass fraction $Y$ increased by 0.05.
In this case, the colour separation $\Delta(F275W-F814W)$ between the two RGBs 
tends to decrease when moving to brighter magnitudes, the opposite behaviour compared to the case 
of metallicity variations.

In addition to these predictions from theory, 
the GC M~2 gives us the opportunity to verify empirically the behaviour of RGBs with different metallicity predicted by the models.

This cluster has been found to host stars characterised by large iron variations, with three main metallicity groups at  [Fe/H] = --1.7, --1.5, and --1.0, respectively \citep{yong14}. While the presence of an intrinsic metallicity spread between the two metal poorer groups (we denote these groups as the main cluster component) is still controversial \citep{lardo16}, several independent studies have confirmed the existence of a small component (accounting for $\sim$1\% of the cluster mass) 
with [Fe/H]=--1 \citep{yong14,miloneM2,lardo16}. Such metal rich stars do not exhibit star-to-star variations in light elements or $s$-process enhancement (e.g. they have P1-like chemical composition; \citealp{yong14}). They are located on a well defined, narrow red sequence which runs parallel to the main RGB body and that can be followed down to the SGB and main sequence, supporting the case for cluster membership \citep{miloneM2}. 

\begin{figure}
\centering
\includegraphics[width=.72\columnwidth]{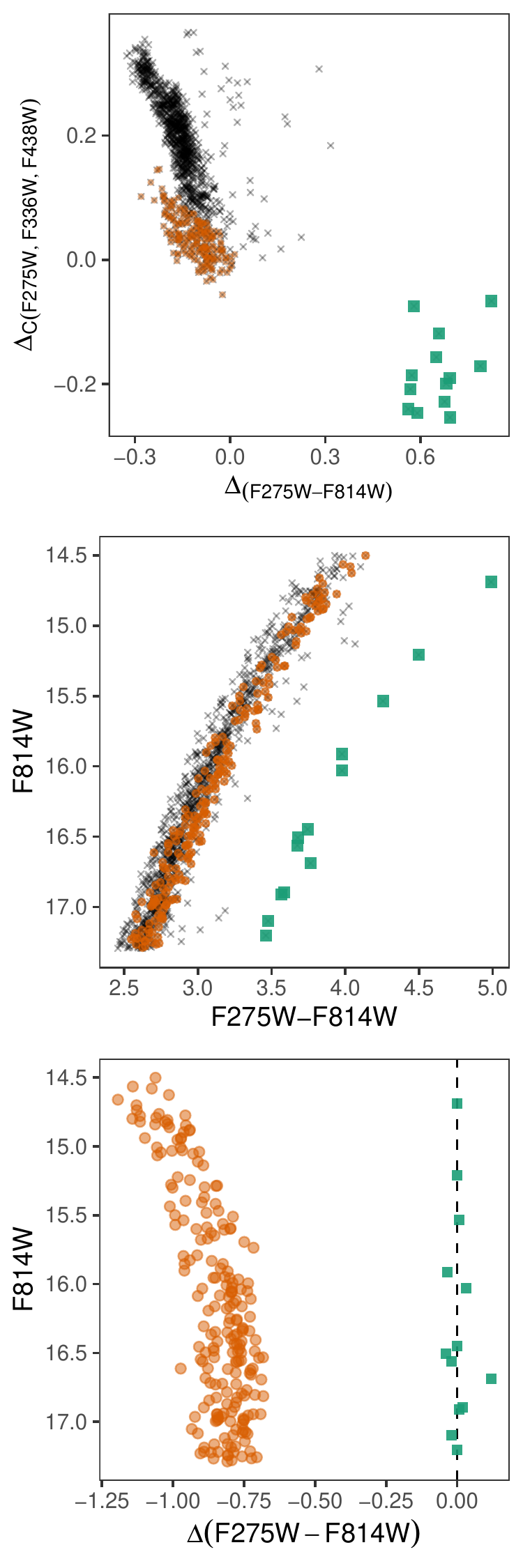} 
\caption{{\sl Top panel:} Chromosome map of M~2. The P1 and P2 stars in the main cluster component are displayed as large orange circles and black crosses, respectively, while the P1 metal rich component is shown as green squares.
{\sl Middle panel:} $F814W$-$(F275W-F814W)$ CMD of the three groups of stars shown 
in the chromosome map.
{\sl Bottom panel:} Similar to the right panel of Fig.~\ref{fig:deltaiso}. The orange circles display the difference 
of the $(F275W-F814W)$ colour between the P1 stars of the main cluster component, and the \lq{verticalised\rq} fit to the CMD of the metal rich component taken as reference (dashed line). The individual metal rich P1 stars are also plotted around the verticalised fit.
\label{fig:m2_new}}
\end{figure}

The top panel of Fig.~\ref{fig:m2_new} displays the chromosome map of the cluster, 
including the metal rich P1 component --that 
comprises a very small fraction of the cluster population, and therefore lies to the red 
of the red fiducials in the diagrams used to calculated the chromosome map-- located at values 
$\Delta_{F275W, F814W}$ larger than 0.5. 
The $F814W$-$(F275W-F814W)$ CMD used to determine the chromosome map is 
shown in the middle panel of 
the same figure, whilst the bottom panel displays the colour 
differences $\Delta(F275W-F814W)$ between stars belonging to the P1 of the main component and the \lq{verticalised\rq} cubic fit to the position of the metal rich P1 stars in the $F814W$-$(F275W-F814W)$ CMD, as a function of $F814W$.
We can see clearly that the $\Delta(F275W-F814W)$ values increase with 
increasing brightness along the RGB, as predicted by theory. 

As a conclusion, the opposite trends of the colour differences $\Delta(F275W-F814W)$ 
with $F814W$ for the case 
of RGBs with different metallicity or different initial helium, provides us with a diagnostic that can be applied to the GCs with extended P1.

\begin{figure}
\centering
\includegraphics[width=.98\columnwidth]{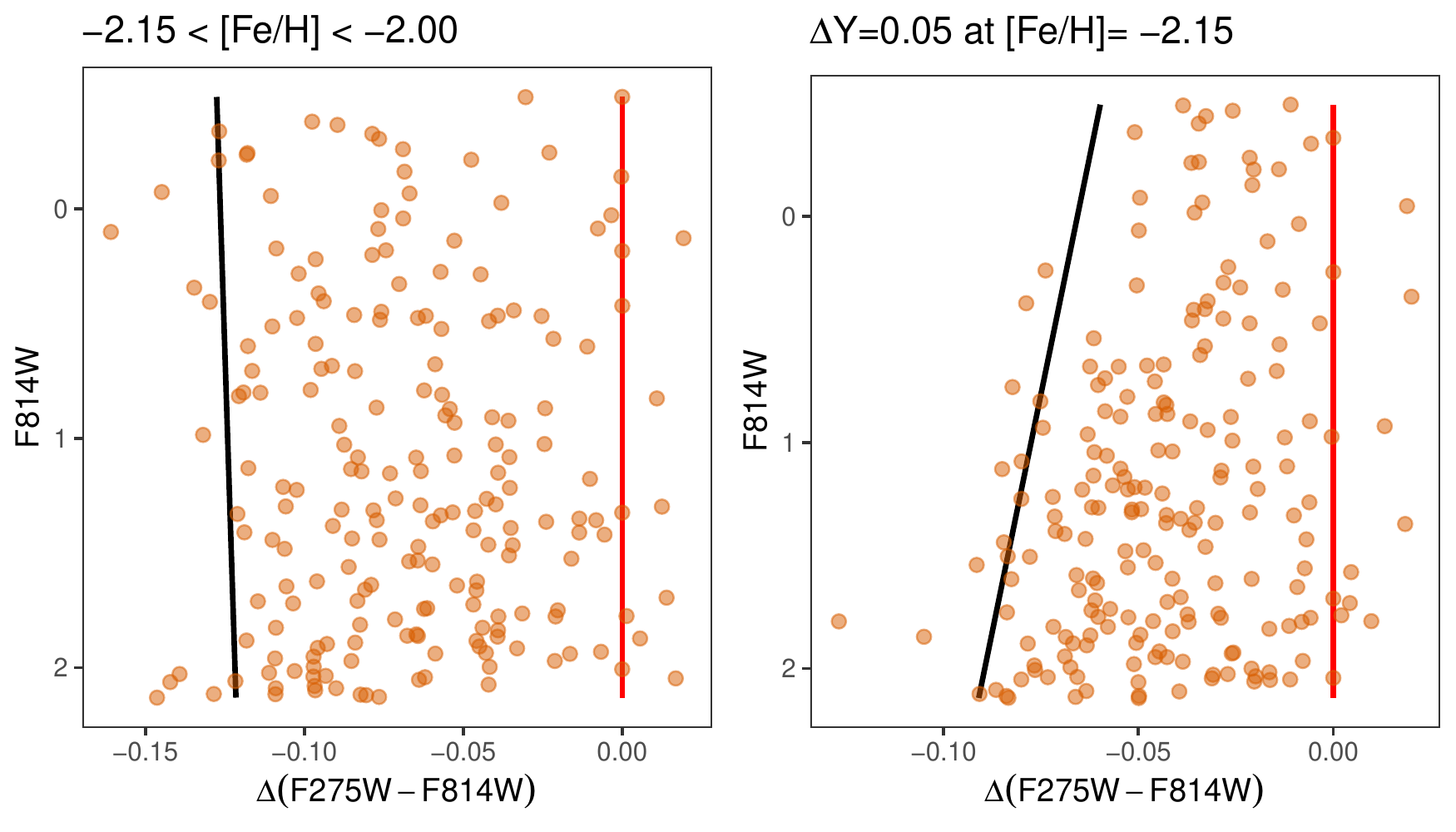} 
\includegraphics[width=.98\columnwidth]{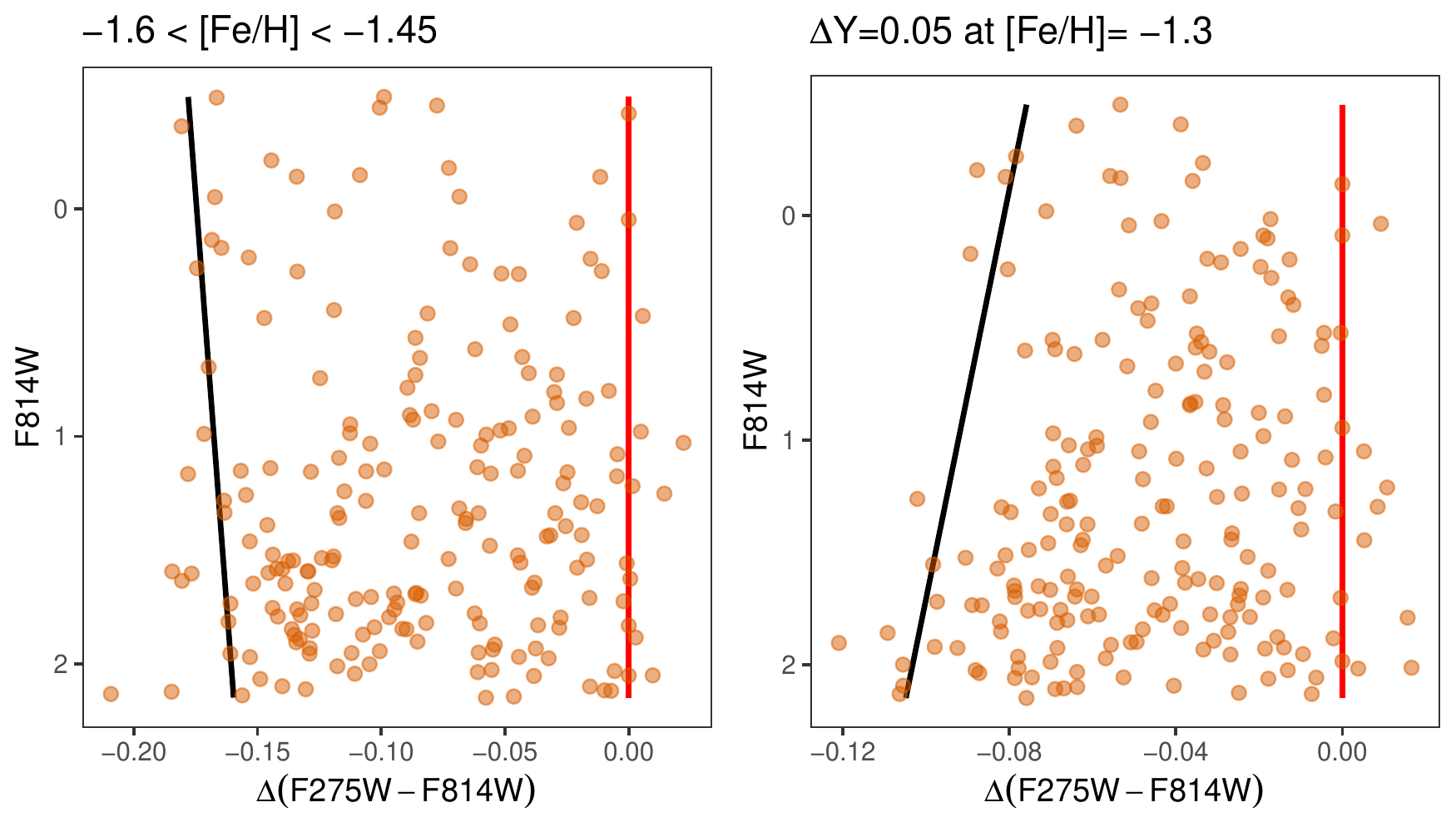} 
\includegraphics[width=.98\columnwidth]{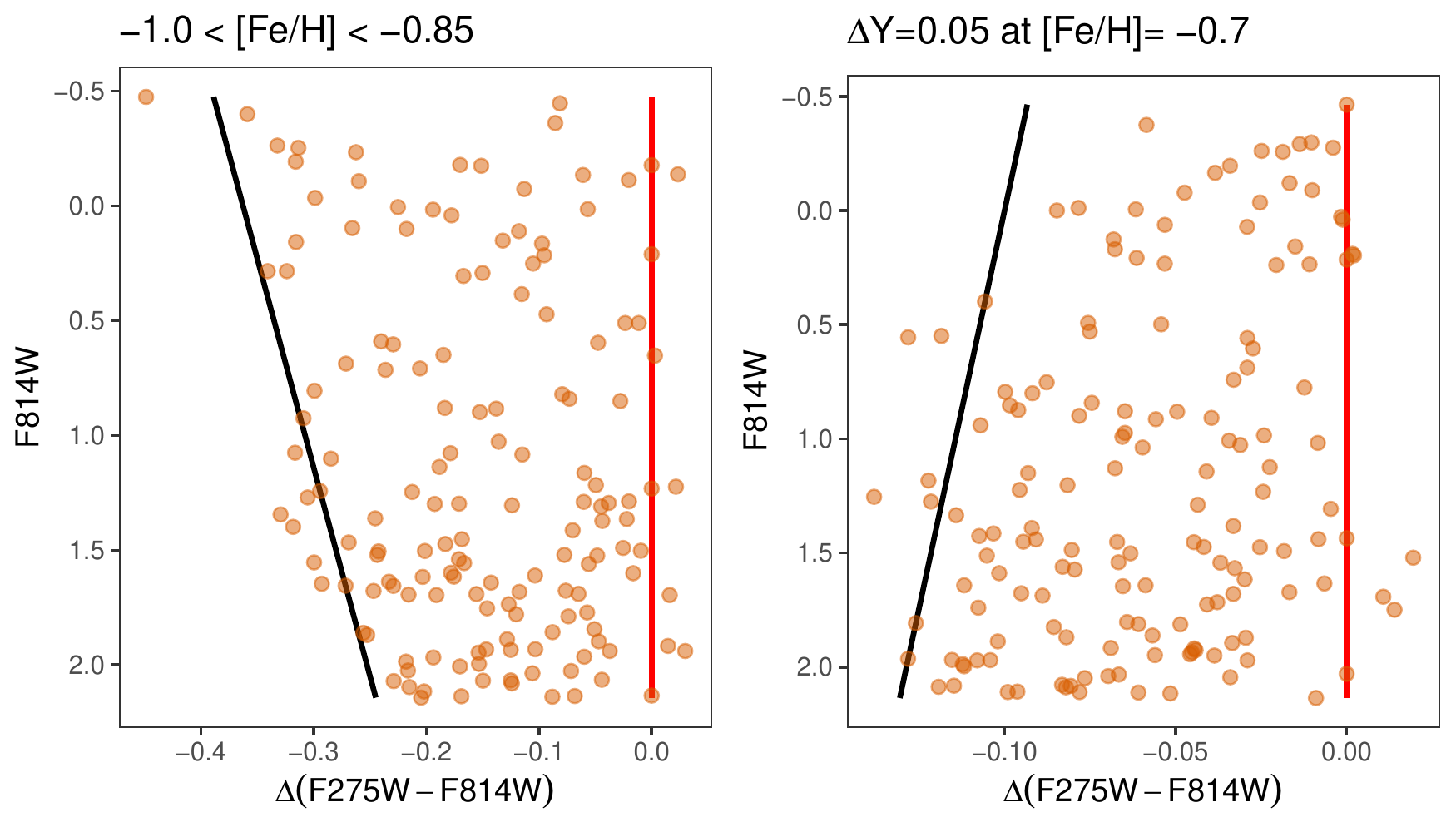} 
\caption{As Fig.~\ref{fig:deltaiso}, but for synthetic samples of P1 stars 
(200 objects) including a 0.15~dex uniform spread of metallicity (left panels), or 
a uniform spread $\Delta Y$=0.05 (right panels).
The top row displays the results for populations with 
[Fe/H] distributed between $-$2.15 and $-$2.0~dex, and for a spread of Helium at 
[Fe/H]=$-$2.15. The middle row shows the case for [Fe/H] between $-$1.6 and $-$1.45~dex, 
and for a spread of helium at [Fe/H]=$-$1.3. The bottom row displays the case 
for [Fe/H] between $-$0.85 and $-1.0$~dex, and for a range of helium at [Fe/H]=$-0.7$.
In each panel the solid vertical lines at $\Delta(F275W-F814W)$=0 are the verticalised red fiducials of the  $F814W$-$(F275W-F814W)$ CMDs of the samples. The other line in each panel correspond to the blue fiducial of the $\Delta(F275W-F814W)$ distribution as a function of $F814W$ (see text for details).
\label{fig:sinObs}}
\end{figure}

To assess how this diagnostic performs in case of populations with a continuous 
distribution of 
helium or metal abundances, we performed a few numerical experiments.
Figure~\ref{fig:sinObs} shows the equivalent of Figs.~\ref{fig:deltaiso} and \ref{fig:deltaisoHe}, but for the $F814W$-$(F275W-F814W)$ CMDs of 
three pairs of synthetic samples with 200 RGB stars each, calculated from 12~Gyr $\alpha$-enhanced isochrones \citep{pietrinferni:06}. This number of stars is a typical average value for the clusters 
in \citet{m17} sample, that we will use for the analysis described below.

The left panels 
of each row display samples with a spread of metallicity (0.15~dex and uniform distribution) while the right panels are samples with a spread of 
$Y$ ($\Delta Y$=0.05 and uniform distribution)\footnote{These ranges of metallicity and $Y$ produce 
widths of the RGB in the $F814W$-$(F275W-F814W)$ CMD, that are consistent with the ranges measured 
by \citet{m17}.}. From top to bottom, the three rows display simulations in order of increasing metallicity regimes. Random Gaussian photometric errors with 1$\sigma$ dispersion typical of  
the cluster photometries used in the next section (average values on the order of $\sim$0.01~mag) are also included in these simulations.

The straight red lines located at $\Delta(F275W-F814W)$=0.0 are the verticalised red fiducials of the  $F814W$-$(F275W-F814W)$ CMDs of the samples. Each red fiducial corresponds to a fit to the 95th percentile (to account for potential outliers due to the photometric error) of the colour distribution as a function of $F814W$.
In the simulations with a metallicity spread, the red fiducials represent the position in the CMD of the populations with the  highest  metallicity, whilst in case of the simulations with a range of $Y$ the red fiducials denote the position of the objects with the lowest $Y$.
The horizontal coordinate of each panel displays the colour difference $\Delta(F275W-F814W)$ between the individual synthetic stars and the corresponding red fiducial at the star $F814W$ magnitude.

The various panels show that, in the case of a metallicity spread, the range of $(F275W-F814W)$ colours spanned by the population increases with increasing $F814W$ luminosity, and this trend steepens 
with increasing mean metallicity. This is evident when considering the 
second straight line displayed in each panel, which is the blue fiducial of the distribution of points in the diagram; it has been derived as a  
linear fit to the 5th percentile of the distribution of the colour differences as 
a function of $F814W$.
In the metal poor regime ([Fe/H] around $-$2) this line is almost vertical, 
and the slope (always in the sense of increasing $\Delta(F275W-F814W)$ with 
increasing $F814W$ luminosity) increases with increasing mean metallicity.
Moreover, for a fixed spread of [Fe/H], the range of $(F275W-F814W)$ colours increases with 
increasing mean metallicity.

The opposite effect can be seen in the case of populations with a range of $Y$.
The range of $\Delta(F275W-F814W)$ values decreases with increasing luminosity, at all metallicities.

The exact value of the slope --not the sign-- of the blue edge of the $\Delta(F275W-F814W)$ distribution as a function of $F814W$, and whether 
the value of the slope is significantly different from zero depends in general on the size of the stellar sample, the photometric error (that in our simulations are set to the typical values of the GC sample we are going to consider in our analysis), the value of the mean metallicity,  
the metallicity (or $Y$) range of the RGB sample, and, very importantly, on the probability distribution of the various metallicities (or $Y$ values) within the adopted range. 

In the case of the choice of parameters of Fig.~\ref{fig:sinObs}, multiple realizations of the simulations show that in both scenarios we can always recover a slope that is significantly different from zero.
We have then varied the probability distribution of metallicity and $Y$ by performing the same simulations described before, but this time 
assuming Gaussian distributions for both metallicity and $Y$. The mean values and 1$\sigma$ spreads were chosen in order to cover, within $\pm$3$\sigma$ from the mean, the same total ranges 
of [Fe/H] and $Y$ as in Fig.~\ref{fig:sinObs}.
Also in this case we formally derive slopes of the blue fiducials of the $\Delta(F275W-F814W)$ distribution as a function of $F814W$ 
with different sign in case of metallicity or $Y$ variations, but in many realizations they were often not significantly different from zero.

The implication of these numerical experiments is that it is possible to use the distribution of P1 RGB stars 
in the $F814W$-$(F275W-F814W)$ CMDs to determine whether a range of metallicity or $Y$ is present, but 
with the available observational data we can find statistically significant results only for certain distributions of metallicity (or $Y$) among the cluster P1 stars. This is similar to the situation with the method employed by \citet{legnardi} based on photometry of subgiants, that can be applied to clusters with P1 and P2 stars well separated along the subgiant branch.

\begin{figure*}
\centering
\includegraphics[width=.92\textwidth]{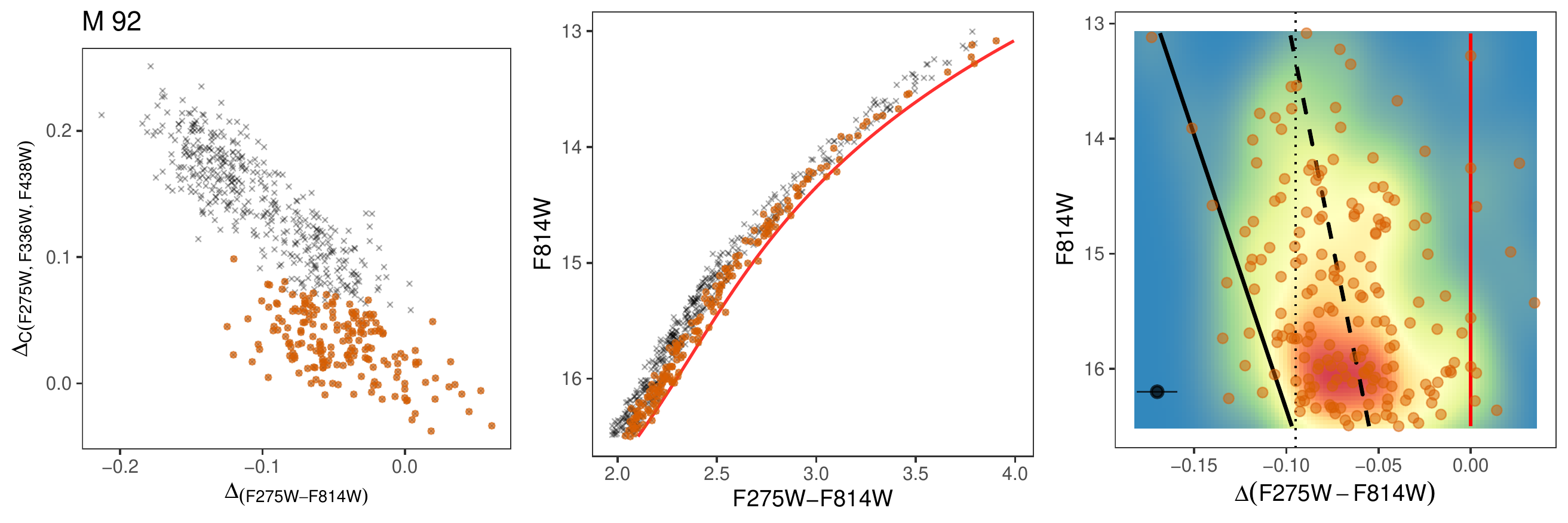} 
\includegraphics[width=.92\textwidth]{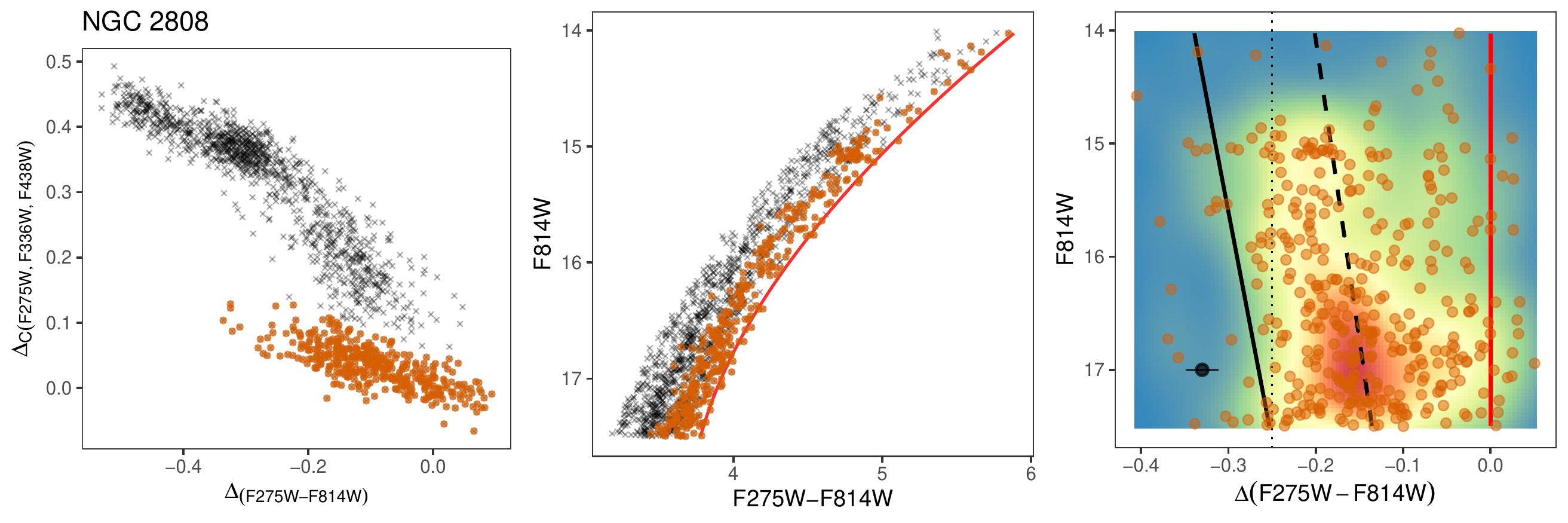} 
\includegraphics[width=.92\textwidth]{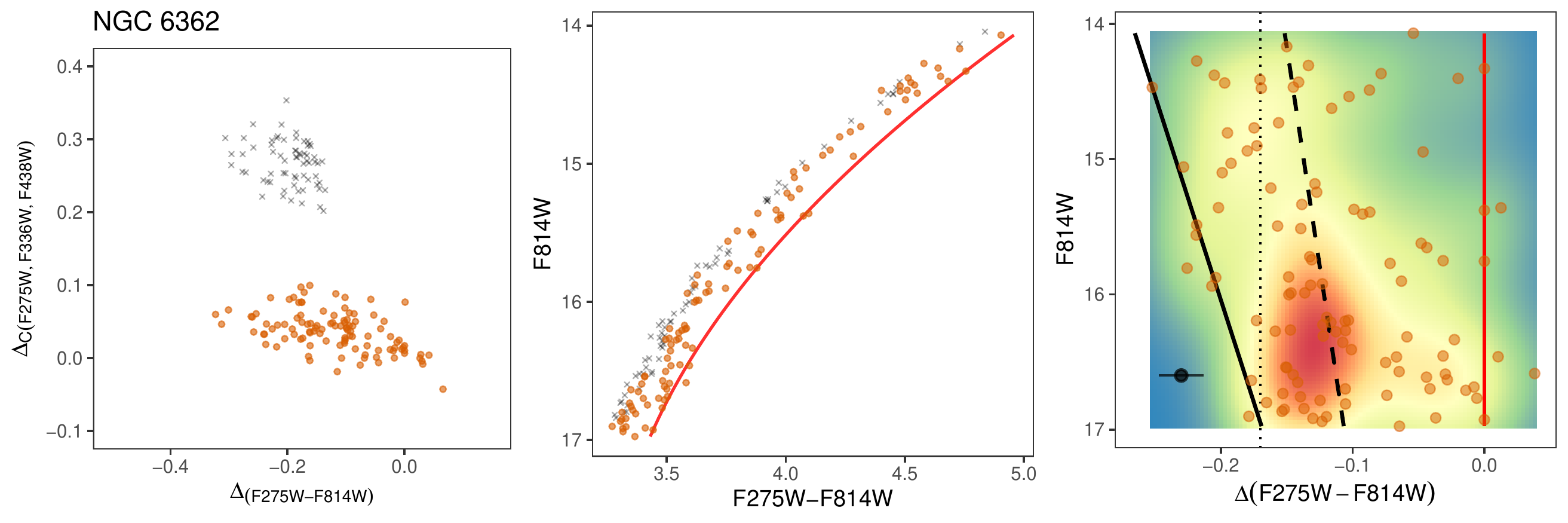} 
\caption{From top to bottom, the three rows of panels refer  
to the clusters M~92, NGC~2808, and NGC~6362 respectively. In each row 
the left panel displays the cluster RGB chromosome map, with small black crosses
denoting P2 stars, and orange circles the P1 population. 
The middle panel displays the 
corresponding $F814W$-$(F275W-F814W)$ CMD, with a 
cubic polynomial fit to the red edge of the P1 sequence (red solid line -- see text).
The faint limit of the CMDs corresponds approximately to 1.5-2.0~mag above the turn off.
The right panel shows the 
difference of the $(F275W-F814W)$ colour of P1 stars 
with respect the \lq{verticalized\rq} fit shown  
in the middle panel (red line). 
The solid black line is a linear best fit to the 5th percentile of the colour differences as a function of $F814W$ (see text for details), corresponding to the blue fiducial of the $\Delta(F275W-F814W)$ distribution as a function of $F814W$, 
while the dotted line denotes a comparison vertical sequence, to emphasise the sign of the slope the blue fiducial in this diagram (see text for details). The dashed black line is the best fit to the 50th percentile of the colour differences as a function of $F814W$.
A colour-coded number density 2D map is also displayed.
The average 1-$\sigma$ $(F275W-F814W)$ photometric error over the $F814W$ covered in the figure is show in the bottom left corner of each panel. 
\label{fig:n2808_new}}
\end{figure*}





\section{The extended P1 of three Galactic globular clusters}
\label{example}

Based on the results of the previous section, we investigate here the behaviour in the $F814W$-$(F275W-F814W)$ CMD 
of RGB stars belonging to three GCs in \citet{m17} sample, 
with well populated RGB photometry in the appropriate filters, and a well extended P1 in their chromosome maps, namely NGC~6341 (M~92), NGC~2808,and NGC~6362 \citep[see][]{m17}, with spectroscopic   
[Fe/H]= --2.35, --1.18, and --1.07, respectively \citep{carrettametal}\footnote{The number of RGB stars in these clusters' photometries range between $\sim$150 (for NGC6362) and $\sim$400 (for NGC~2808).}.
Unfortunately we cannot employ in our analysis the GC NGC~3201 --whose P1 RGB stars display a range of metallicities according to the spectroscopic analysis by \citet{marino:19}-- because the available photometry does not have enough stars along the RGB to perform  our analysis (see the  previous section).

For each cluster we have determined first the chromosome map, to identify their P1 stars. 
Given that we are interested in high-precision photometry of cluster members, we retain for our analysis only stars with membership probability greater than $>95\%$. Additionally, only stars with relatively small photometric uncertainties ($<$0.05, 0.04, 0.03, 0.02, 0.01~mag in F275W, F336W, F438W, F606W and F814W  respectively) have been selected to produce the chromosome maps. 

Cluster members along the RGB have been identified by performing a multivariate analysis on the photometric data through a Gaussian mixture model algorithm (GMM; \citealp{gmm}). For each giant star in a given cluster photometry, we have computed colour (and colour-indices) from all possible combination of the magnitudes available in the \citet{nardiello19} catalogues --including $(F275W-F814W)$ and $C_{F275W, F336W, F438W}$. Member stars have then been selected using the constraint that similar objects (e.g. stars belonging to the cluster) lie on a well-defined RGB sequence in the multi-dimensional space defined by all the possible CMD combinations. In such a way, stars which are likely contaminants (field stars and/or binaries) are easily spotted by the algorithm because they are scattered around when using different colour (or colour index) combinations to create the CMD. 

Once the cluster RGB stars have been selected, we have followed the procedure described in Sect.~\ref{method} to derive the chromosome maps shown in Fig.~\ref{fig:n2808_new}.
The P1 stars have been then identified according to their position in the chromosome map \citep[following][]{m17} and they are plotted as orange circles in figure~\ref{fig:n2808_new}.

The left hand panels of Fig~\ref{fig:n2808_new} show the chromosome maps of (from top to bottom) M~92, NGC~2808, and NGC~6362 respectively. 
From a visual comparison they are very similar to the ones published by \citet{m17}. Also, the number ratios of P1 stars to the total number of RGB stars we obtained for these clusters are  consistent with the values listed in Table~2 of \citet{m17}.

The middle panels of Fig.~\ref{fig:n2808_new} show the position of the RGB stars of the chromosome maps in the $F814W$-$(F275W-F814W)$ CMDs. The redder envelope of the P1 population in these CMDs has been derived by fitting the 95th percentile (to account for potential outliers) of the 
colour distribution as a function of $F814W$ with a third order polynomial. 
This red fiducial line is shown in red in the middle panels of the same figures, and corresponds 
to the location of P1 stars with either the highest metallicity --if metallicity varies 
along the extended P1 sequence in the chromosome maps-- or the lowest helium abundance --if helium changes along the P1 sequence.

The $\Delta(F275W-F814W)$ colour differences between the individual RGB stars and the fiducial lines in the middle panels are plotted in the right hand panels of Fig.~\ref{fig:n2808_new}. The red solid lines denote the verticalised red fiducials described before, whilst the black solid lines are linear fits to the 5th percentile of the $\Delta(F275W-F814W)$ distribution as a function of $F814W$, and 
represent the blue edge, or blue fiducial, of the distribution of $\Delta(F275W-F814W)$ values.
We have verified that for each cluster the sign of the slope of this blue fiducial does not change if we fit alternatively the 10th or even the 15th percentile of the colour distribution, to be more conservative regarding the presence of unresolved binaries and blue 
stragglers progeny 
(that have been already largely filtered out by our procedure to determine the chromosome maps) which are predicted to be 
located mainly at the bluer end of the extended P1 sequence in the chromosome maps 
\citep[see][]{martins, marino:19}.

We find in all clusters that the range of $\Delta(F275W-F814W)$ values increase  with increasing $F814W$ brightness; this is a clear signature for the presence of a range of metal abundances among their P1 stars (see Fig.~\ref{fig:sinObs}). Errors on the magnitudes show a decreasing trend with increasing luminosity, thus the photometric error does not contribute to the broadening of the RGB with luminosity.
The lines denoting the median of the colour differences as a function of $F814W$, and the number density of points in these diagrams also show  
in all three clusters a general trend consistent with the sign of the slope of the blue fiducial. The absolute values of the slope of the blue fiducial in all three clusters are significantly different from zero at more than 95\% confidence level.
A range of helium as the 
main driver of the observed range of $(F275W-F814W)$ colours among P1 stars seems to be ruled out. 

\section{Summary and discussion}
\label{discussion}

We have exploited $HST$ near-UV and optical photometry of RGB stars in the Galactic GCs 
M~92, NGC~2808, and NGC~6362, to determine whether metal or helium abundance variations are the cause of the extended P1 sequences in their chromosome maps.
The cluster NGC~6362 is in common with \citet{legnardi} study, and we find for this cluster and all others in our sample that a range of metal abundance does exist among their P1 stars.
This confirms the results of \citet{legnardi} based on the analysis of subgiant branch stars in NGC~6362 and NGC~6838, and solidifies the case 
for the existence of unexpected variations of metal abundances in most globular clusters.

\begin{figure}
\centering
\includegraphics[width=.479\textwidth]{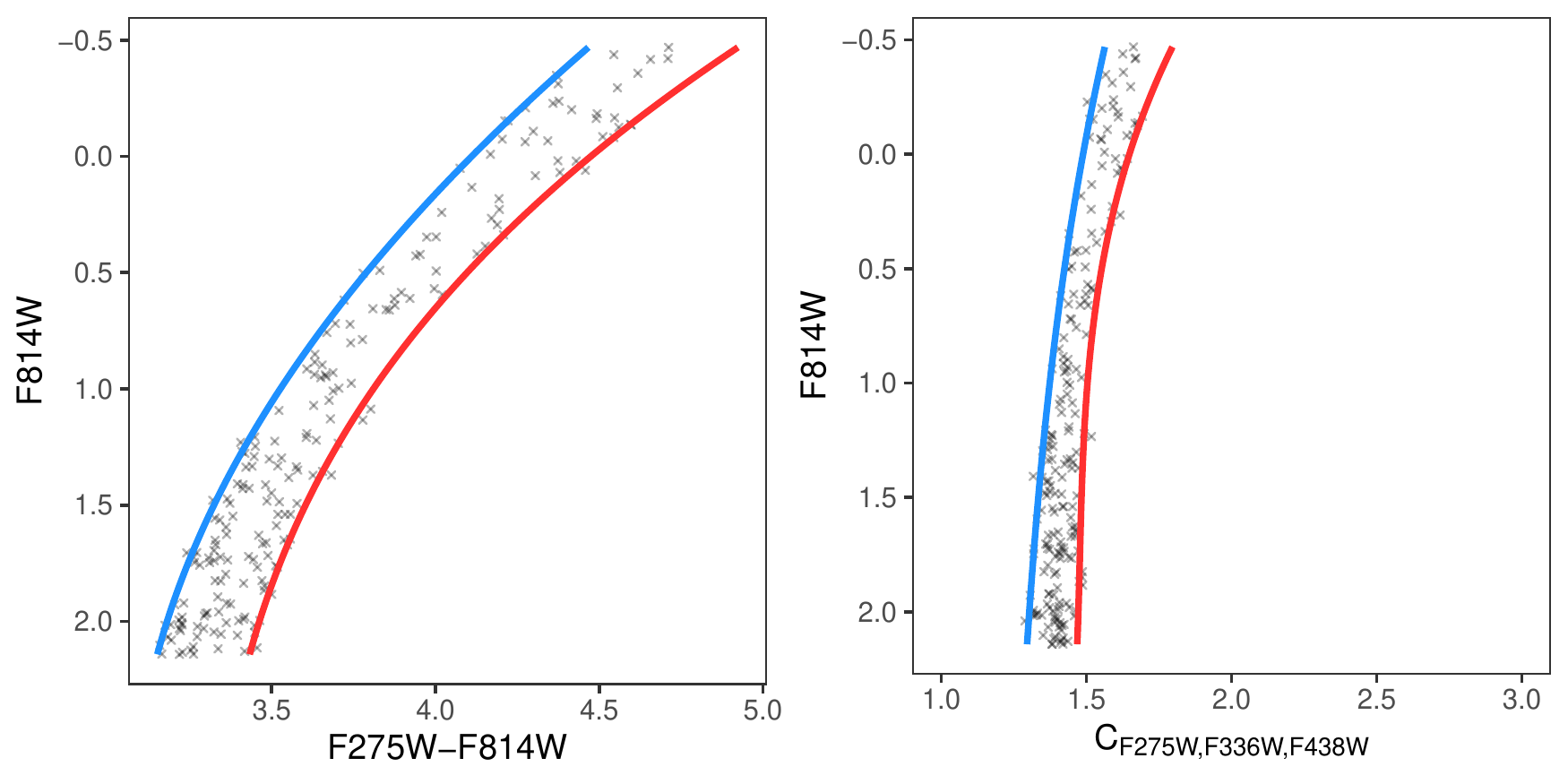} 
\caption{{\em From left to right:} the $F814W$-$(F275W-F814W)$ and $F814W$-C$_{(F275W,F336W,F438W)}$ 
diagrams for a synthetic sample with --1 $<$ [Fe/H] $<$ --0.85) like the one shown in Fig.~\ref{fig:sinObs}.
The blue and red fiducial lines shown in each panel as solid lines (determined as described in Sect.~\ref{method}) 
highlight the different widths of the RGB in the two diagrams. 
The width (in magnitudes) of the horizontal axis of the two figures is the same.
\label{fig:p1_map}}
\end{figure}

A range of metallicity does not only explain the P1 extension in the $(F275W-F814W)$ colour, but also the associated smaller range of the $C_{F275W, F336W, F438W}$ pseudocolour covered by P1 stars in the 
chromosome maps. This is shown in 
Fig.~\ref{fig:p1_map}, which displays the $F814W$-$(F275W-F814W)$ and the $F814W$-$C_{F275W, F336W, F438W}$ diagrams of a synthetic population with --1 $<$ [Fe/H] $<$ --0.85, like the one shown in Fig.~\ref{fig:sinObs}.

Assuming that a metallicity spread --as found for NGC~6362 and NGC~6838-- is the cause of the P1 extension in the chromosome maps of all clusters in \citet{m17} sample, \citet{legnardi} determined the correspondent range of [Fe/H] 
--a proxy for a range of total metallicity in P1 stars-- for the sample studied by \citet{m17}, comparing the width of the RGB ($W_{F275W,F814W}$) taken two $F814W$ magnitudes above the main sequence turn off tabulated by \citet{m17} with theoretical isochrones.
They derived internal variations ranging from less than 0.05 to 0.30~dex, that mildly correlate with cluster mass and metallicity.
Our study corroborates these conclusions, by directly 
establishing the existence of a metallicity spread in a slightly larger sample of clusters distributed over a wider [Fe/H] range compared to \citet{legnardi} study, using an independent method.

Figure~\ref{fig:n2808_new} also shows that 
the P2 distribution in the $F814W$-$(F275W-F814W)$ CMDs of NGC~6362 appears clustered around the blue edge of the P1 distribution, whilst for both M~92 and NGC~2808 the red 
edge of the P2 distribution is close to the red edge of the P1 one.
Also, in case of M~92 the width of the P2 sequence is 
comparable to the width of the P1 one, whilst it is larger in NGC~2808.
This has implications for the metallicity distribution of P2 stars in these clusters, 
if we consider the analysis by \citet{lsb18} \citep[see also][]{He18} who have shown that at fixed metallicity the $(F275W-F814W)$ colour of RGB P2 stars becomes bluer when the extension of the light element anticorrelations (due mainly to the decrease of oxygen) and $Y$ increase.

The red edge of the P2 CMD corresponds to the P2 stars 
with the smaller values of the light element abundance anticorrelations and 
helium enhancement, hence its position with respect to the P1 colour distribution 
is likely a fair indicator of the metallicity out of which these P2 stars have formed.
In case of NGC~6362 these P2 stars have formed with the same [Fe/H] of the metal poorer P1 population, while in the other clusters they formed with a [Fe/H] closer to the metal richer P1 component.
Also, given that at fixed [Fe/H] the $(F275W-F814W)$ colour of RGB P2 stars increases 
with increasing $Y$ and decreasing oxygen, a comparable or narrower width of the P2 
sequence compared to the P1 counterpart suggest a smaller 
[Fe/H] range in the P2 cluster population.
These conclusions echo those by \citet{legnardi}, based on the distribution of the P1 and P2 stars in the chromosome maps, and highlight the emerging ever increasing complexity of the chemical makeup of GCs.

The existence of a metallicity spread amongst P1 stars and a different distribution of the total metallicity in P1 and P2 populations might have implications for the determination of the relative He abundances, based on the comparison of theoretical and observed
colours of P1 and P2 RGB stars \citep{He18}, that need to be assessed. We also need to investigate the 
impact of both helium and total metallicity variations on the interpretation of the horizontal branch morphologies of individual clusters, using synthetic horizontal branch modelling 
\citep[see, e.g.,][]{dale13, csp14, scp, tailo19}.

Finally, it is also important to further validate these results about metallicity spreads with dedicated high-resolution spectroscopic analyses.
So far, as already mentioned, spectroscopy of NGC~2808 P1 stars distributed along the extended sequence in the chromosome map has not disclosed any metallicity spread, and 
for example the spectroscopic analysis of a sizable sample of P1 RGB stars in NGC~6362 by \citet{muc16} did not reveal any statistically significant spread. 
This lack of detection might be due to measurements not sufficiently precise compared to the range of metal abundances in these two clusters, that are  
equal to $\sim$0.1~dex and $\sim$0.05~dex for NGC~2808 and NGC~6362 respectively \citep[see][]{legnardi}. Moreover, in case of NGC~6362 it is possible that P1 stars on 
the blue side of the chromosome map have not been sampled by the observations.
In the case of M~92 \citet{langer} found a spread of 0.18~dex (consistent with 
$\sim$~0.15~dex estimated by \citealp{legnardi} for P1 stars) in iron peak elements among three bright red giants, but a recent analysis by \citet{mesz} does not find any significant metallicity spread in this cluster.


\begin{acknowledgements}
The authors would like to thank Ivan Cabrera-Ziri for 
useful comments  and discussions.
M. Salaris acknowledges support from The Science and Technology Facilities Council Consolidated Grant ST/V00087X/1.
C. Lardo acknowledges funding from Ministero dell'Università e della Ricerca through the Programme {\em Rita Levi Montalcini} (grantPGR18YRML1).
Isochrones and photometry used in this study are available at \url{http://basti.oa-teramo.inaf.it/index.html} and 
\url{https://archive.stsci.edu/prepds/hugs/}, respectively.
\end{acknowledgements}

%
\bibliographystyle{aa} 
\bibliography{paperbib} 

\end{document}